# Revisiting Fourier-Based Reflection Spectrum Measurement in Coherence Scanning Interferometry: Overlooked Impact of PCOR in High-NA Systems

CHENG CHEN,[1] SOTERO ORDONES,[1] JEREMY COUPLAND,[2] AND RONG SU[1]

[1] *Shanghai Institute of Optics and Fine Mechanics, Chinese Academy of Sciences, Shanghai 201800, China*
[2] *Wolfson School of Mechanical, Electrical and Manufacturing Engineering, Loughborough University, Loughborough LE11 3TU, UK*
\* *surong@siom.ac.cn*

**Abstract:** Recent advancements have extended the capabilities of coherence scanning interferometry (CSI) beyond surface topography measurement to reflectivity spectrum imaging. It is commonly accepted that the one-dimensional(1-D) Fourier magnitude of a pixel-wise CSI signal is proportional to the product of the sample's absolute reflection coefficient spectrum and the light source spectrum. However, this assumption holds true only for low numerical aperture (NA) CSI systems. In this work, we establish a generalized mathematical framework that represents the CSI signal a matrix multiplication of the spectrum measurement matrix with the element-wise product of the reflection spectrum and the light source spectrum, assuming spectral independence from the incident angles. Considering the light source spectrum and wavenumber channel summation, we utilize a reflection spectrum measurement matrix and its conditioned counterpart to linearly relate the CSI signal to the absolute reflection spectrum. Through simulations, we quantitatively analyze the impacts of various error sources, including NA inaccuracy, pupil apodization, height reconstruction errors, spectral inaccuracies of the light source, and phase change on reflection (PCOR) from the sample, on the reconstruction performance of the absolute reflection coefficient spectrum. It is found that the PCOR is the dominant factor influencing the reconstruction of the reflection coefficient spectrum.

## 1. Introduction

Coherence scanning interferometry (CSI) can achieve surface roughness and 3D structural measurement at the microscopic scale. Its full-field imaging capability enables the assessment of surface topography and roughness with both high lateral resolution and sub-nanometer axial noise levels across millimeter-scale fields of view [1-3].

A typical CSI setup includes a broadband light source, Köhler illumination optics, an interference objective, and a mechanical scanner [1]. The Köhler illumination optics projects the extended light source into the interference objective's pupil, creating uniform incoherent illumination at the sample plane. The mechanical scanner provides a smooth continuous scan of the objective in the z-direction. During the scan, a sequence of interference fringes is recorded at each scanning positions. Techniques, such as envelope detection, Fourier domain analysis and correlation between the measured and reference fringes, are typically used to recover topographic information [4-5].

Recent advancements in hardware and image processing techniques have extended CSI capabilities to include reflectivity spectra imaging. J. L. Beverage et. al [6] proposed to use a multi-color-channel camera or a multi-color LED light source to derive colorimetric characteristics of the sample. But spectral resolution is limited by the hardware such as the multi-color-channel camera or the multi-color light source, and there is a trading off CSI's standard performance in exchange for the added colorimetric capability.

On the other hand, quantitative reconstruction of reflectivity spectra can also be achieved through Fourier analysis of CSI fringes. Xu et. al [7], and Arnaud Dubois et .al [8] have applied the one-dimensional (1-D) Fourier transform to pixel-wise fringe signals to attain high spectroscopic resolution reflection or absorption spectra in optical coherence tomography, which has the same basic functional principles as CSI. However, R. Claveau et. al [9] and P. Lehmann et. al [10] have demonstrated that the Fourier magnitude spectrum significantly deviates from the light source's spectrum, when measuring a flat mirror of constant reflectivity across different wavelengths, with a CSI whose numerical aperture (NA) ≥ 0.3. These discrepancies have been confirmed through both simulations and experiments, as noted in [10]: "The light source is a blue LED with a center wavelength of 460 nm, resulting in a maximum effective wavelength of 644 nm for NA = 0.55 and 1055 nm for NA = 0.9."

R. Claveau and his colleagues [11-12] have interpreted the Fourier magnitude of a pixel-wise signal as the product of the spectral transfer function of the CSI system and the square root of the sample's reflectivity spectrum (i.e., absolute reflection coefficient spectrum) in a high NA CSI. The spectral transfer function is calibrated using the Fourier magnitude calculated from the measured fringe signals of a reference sample-such as silicon wafer-with a known reflectivity spectrum. With this method, the parallel mapping of local spectral and topographic information of gold on an aluminum sample and colored stainless steel have been achieved [11-12]. However, the illumination aperture is intentionally closed in the experiments, indicating a small illumination and imaging NA for flat samples. More importantly, the mutual coupling between magnitude and phase in reflection coefficient spectrum reconstruction remains unrevealed.

In this work, we present a framework to analyze the sample's reflection spectrum reconstruction in CSI. We begin by reviewing the widely accepted 1-D signal model, which accounts for both spatial and spectral coherence, and then simplify this model into a linear algebraic representation of the CSI signal. Several measurement matrices for the reflection spectrum reconstruction are used to characterize the CSI system, and the error sources are quantitatively analyzed through simulations.

## 2. Signal model for a high NA CSI

### 2.1 1-D signal model for a high NA CSI

In general, the 1-D Fourier analysis of the pixel-wise CSI signal starts with NA equal to zero [7,9]. Here, we use the 1-D signal model that takes both the spatial and temporal coherence into consideration[13-15] .The CSI signal at pixel-level is an integral over all points in the pupil plane and over all wavelengths for the ray bundle contributions[16-17].

$$I(z) = \int_0^\infty \int_{\psi_1}^{\psi_2} g(k,\psi,z) P(\psi) s(k) \sin\psi \cos\psi \, d\psi \, dk , \qquad (1)$$

where $z$ denotes the scanning position, $k = 1/\lambda$ is the wave number at an illumination wavelength $\lambda$, $\psi$ is the incident angle within the range of $[\psi_1, \psi_2]$, $s(k)$ denotes the optical spectrum distribution of the light source, $P(\psi)$ is the intensity distribution in the pupil plane of the objective, with the implicit assumption of rotationally symmetric pupil apodization, and $g(k,\psi,z)$ denotes the interference signal for a single ray bundle at incident angle $\psi$ [18-19],

$$g(k,\psi,z) = 2\Re\{r_r(k,\psi)^* r_s(k,\psi) \exp[j4\pi k(h_s - z)\cos\psi + j\Delta(k,\psi)]\} , \qquad (2)$$

where $\Re$ is the taking the real part operation, $r_r$ denotes the complex reference reflection coefficient, including both the beam splitter and the reference mirror, $r_s$ denotes the sample's complex reflection coefficient, including, e.g., the transmission of the beam splitter, $h_s$ is the sample surface height, * is the complex conjugate operation, and $\Delta$ is an additional phase term associated with the chromatic dispersion of the CSI system. The background signal is not taken into consideration here.

Equation (2) can be re-written as

$$g(k,\psi,z) = 2|r_r(k,\psi)||r_s(k,\psi)|\cos[4\pi k(h_s - z)\cos\psi + \Delta(k,\psi) + \Delta_s(k,\psi)]\,, \quad (3)$$

where $\phi_s(k,\psi)$ is the phase of the sample's reflection coefficient—often referred to as the phase change on reflection (PCOR)— at different wavenumbers and incident angles, $\phi_r(k,\psi)$ is the corresponding phase for the reference path, and $\Delta_s(k,\psi) = \phi_s(k,\psi) - \phi_r(k,\psi)$ is the phase difference between them. Usually, the reference arm ends in a mirror that introduces its own PCOR, so the interference depends not on the individual phases but on their difference: the relative PCOR $\Delta_s(k,\psi)$ between the sample and reference paths.

Since the reflection coefficients of a sample depend on its polarization properties, it is important to account for the polarization state in a CSI system. In this manuscript, a non-polarized CSI setup is considered. Given that the measured non-transparent samples are isotropic with respect to polarization, the incoherent superposition model can accommodate polarization effects by averaging the sample's reflection coefficients under s- and p-polarizations [17], provided that the polarization evolution is neglected when through a high NA microscope objective [20].

## 2.2 Assumption of spectral independence from the incident angles

When the measured sample is a flat surface coated with pure metallic materials such as silver (Ag), gold (Au), and copper (Cu), the sample's complex reflection coefficient can be calculated using the Fresnel equation as follows [21],

$$r^s = \frac{n_1\cos\psi - n_2\sqrt{1-\left(\frac{n_1}{n_2}\sin\psi\right)^2}}{n_1\cos\psi + n_2\sqrt{1-\left(\frac{n_1}{n_2}\sin\psi\right)^2}}, \quad (4)$$

and

$$r^p = \frac{n_2\cos\psi - n_1\sqrt{1-\left(\frac{n_1}{n_2}\sin\psi\right)^2}}{n_2\cos\psi + n_1\sqrt{1-\left(\frac{n_1}{n_2}\sin\psi\right)^2}}, \quad (5)$$

where $r^s$ and $r^p$ denote the complex reflection coefficients for the sample under $s-$ polarized and $p-$polarized light, respectively, while $n_1$ and $n_2$ denote the refractive indices of the environment and the material.

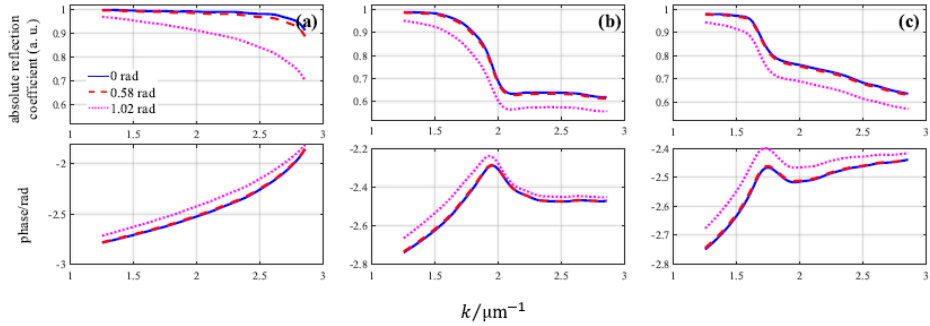

Fig. 1. Complex reflection spectra for various materials:(a) Ag, (b) Au, and (c) Cu.

Figure 1 plots the complex reflection coefficients —averaged over s- and p-polarizations— for three non-transparent materials (Ag, Au, and Cu) across wavenumbers and incident angles.

Minimal difference exists between normal incidence and angles≤ 0.58 rad (NA=0.55) in both the absolute reflection coefficient spectrum and its phase spectrum. Significant deviations emerge at oblique incidence angles up to 1.02 radians (NA=0.85) when compared to normal incidence.

Given the marginal angular dependence below NA=0.55, we restrict CSI operation to this regime where spectral responses can be considered incident-angle-independent. Consequently, Eq. (3) simplifies to:

$$g(k,\psi,z) = 2|r_r(k,\psi)||r_s(k,\psi)|\cos[4\pi k(h_s - z)\cos\psi + \Delta(k,\psi) + \Delta_s(k)], \quad (6)$$

where the complex reflection coefficient of reference path $r_r(k)$, and the sample's complex reflection coefficient $r_s(k)$ are incident angle-independent.

### 2.3 Linear algebraic representation of CSI signal

Before delving deeper, we introduce a notation in linear algebra. Typically, a "2-D matrix" is represented by bold, italic, uppercase letters; a "1-D vector" by bold, italic, lowercase letters; and a scalar value by italic, lowercase letters.

In discrete formulation, the light source's spectral power distribution is sampled at $n$ equally spaced wavenumbers $k_1, k_2, \ldots, k_n$ across the spectral bandwidth. The CSI signal is an incoherent, weighted sum of monochromatic interferometric contributions at these discrete wavenumbers, with each weight equal to the source's spectral intensity at the corresponding wavenumber. When reflection coefficients are angle-independent as in Eq. (6), the sample-dependent term factorises out of every monochromatic signal. Consequently, the weights become the product of the source's spectral spectrum and the magnitude of the reflection spectrum.

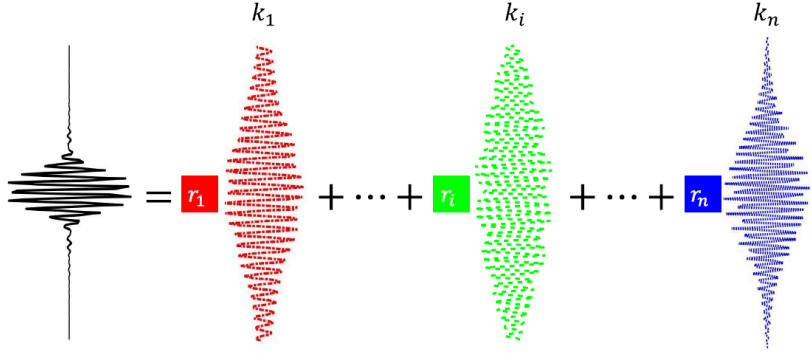

Fig. 2. Linear interpretation of the CSI signal.

Accordingly, the discrete expression in Eq. (1) simplifies to:

$$\boldsymbol{y} = \boldsymbol{A}\boldsymbol{r}, \quad (7)$$

where $\boldsymbol{y}$ is the $m$-point interferometric signal vector, $m$ is the number of the scanning positions, $\boldsymbol{A}$ is the spectrum measurement matrix, and $\boldsymbol{r} = |\boldsymbol{r}_r| \odot |\boldsymbol{r}_s| \odot \boldsymbol{s}$ is the real-valued spectrum vector. Here, $|\boldsymbol{r}_r|$ and $|\boldsymbol{r}_s|$ represent the discrete vectors of the absolute reflection coefficients for the reference part and the sample, respectively, while $\boldsymbol{s}$ denotes the vector of the spectral distribution of the light source. The operator $\odot$ represents element-wise multiplication. The $(i,j)$th entry $a_{ij}$ in the spectrum measurement matrix $\boldsymbol{A}$ is given as:

$$a_{ij} = 2 \int_{\psi_1}^{\psi_2} P(\psi) \cos[4\pi k_j (h_s - z_i)\cos\psi + \Delta(k_j, \psi) + \Delta_s(k_j)] d\psi, \tag{8}$$

where subscript $i$ denotes the $i$th scanning position, subscript $j$ denotes the $j$th wavenumber, and $\Delta_s$ denotes the angle-independent phase difference between the complex reflection coefficients of the reference part and the sample part.

Equation (7) straightforwardly follows from interpreting the CSI signal as the sum of monochromatic interferometric signals at the discrete wavenumbers $k_1, k_2, \ldots, k_n$, each weighted by the corresponding element of the spectrum vector $r$, as shown in Fig. 2. Each column $[a_{1i}, \ldots, a_{mi}]^T$ of the spectrum measurement matrix $A$ is the monochromatic interferometric signal vector, obtained when both the source intensity and the sample's reflection magnitude are set to unity.

## 3. Measurement matrices for reflection spectrum reconstruction in CSI

The inherent nature of Eq. (7) can be understood from different perspectives through numerical analysis of the spectrum measurement matrix. In the following simulations, we utilize a Gaussian distributed light source with a center wavelength of approximately 0.57 μm and a full-width at half-maximum (FWHM) of about 0.12 μm as shown in Fig. 3. The spectral bandwidth ranges from [1.3628 μm$^{-1}$, 2.2375 μm$^{-1}$], encompassing 45 discrete wavenumbers. Furthermore, the intensity values at the two boundary wavenumbers are set to be no smaller than 0.1% of the maximum intensity. The scanning interval is set as 0.578/8 μm, and the number of scanning positions is 345. The sample surface height $h_s$ is set to 0. The chromatic dispersion of CSI system $\Delta(k, \psi)$ is neglected here, and a uniform pupil distribution is assumed. And the material Au is measured here. Unless otherwise specified, these setting remain unchanged in following sections.

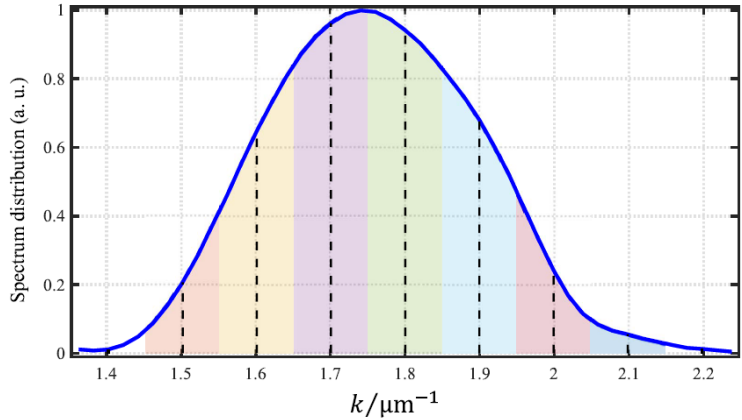

Fig. 3 Spectral distribution of the light source in following simulations.

### 3.1 Inherent nature of spectrum measurement matrix

In Fig. 4, an example of the spectrum measurement matrix $A$ is shown, where each column of the matrix denotes the monochromatic interferometric signal at a corresponding wavenumber. This not only lays the foundation of our simulations, but also offers a computationally efficient method — approximately two orders of magnitude faster than direct integration— to simulate the CSI signal.

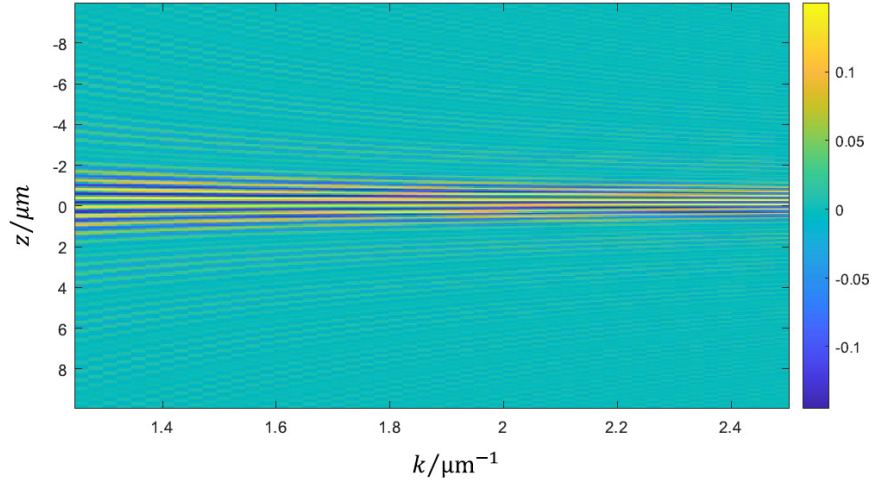

Fig. 4. Illustration of the spectrum measurement matrix $A$ at NA= 0.55.

When analyzing the linear problem in Eq. (7), the first step is to determine whether a unique solution exists. Figure Fig. 5 shows the normalized singular values of the spectrum measurement matrix $A$ for several NAs. Because the smallest singular value is strictly positive in every case, $A$ is full rank, so Eq. (7) admits a unique least-squares solution. In principle, therefore, the unknown spectrum vector $r$ can be recovered with the pseudoinverse of $A$, provided $A$ itself is known exactly. Numerical stability, however, is governed by the condition number. As Fig. 5 also indicates, condition number lies between about 50 and 100 for the NAs studied. Such a range means that small errors in $A$ or in the measurement vector $y$ can be amplified by up to two orders of magnitude in the solution, so Eq. (7) is moderately ill-conditioned despite being perfectly solvable.

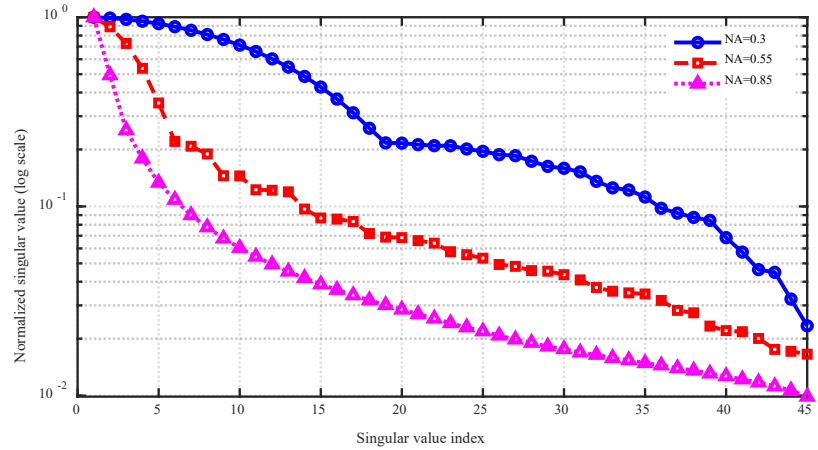

Fig. 5. Normalized singular values of the spectrum measurement matrix at different NAs.

Re-examining the spectrum-measurement matrix $A$ from a column-orthogonality perspective clarifies the link between rank, singular values, and conditioning. Full rank merely guarantees that no column is an exact linear combination of the others; it does not imply strong independence. A steep drop in the singular-value spectrum of the matrix therefore signals that, while redundancy is absent, many columns are still highly correlated.

To visualize where those correlations come from, we move to the spatial-frequency domain. Figure 6 (a) plots three representative columns of the spectrum measurement matrix $\boldsymbol{A}$ (the 4th, 23rd, and 45th columns in Fig. 4), each corresponding to the interferometric signal under a single wavenumber. Taking the 1-D Fourier transform of a column yields its 1-D monochromatic transfer function (TF), illustrated in Fig. 6 (b) and (c). It is straightforward that the support of TFs at adjacent wavenumbers overlaps in nonzero NA cases, as the monochromatic TF covers a substantial range in the Fourier domain, ranging from $[2k\sqrt{1-\mathrm{NA}^2}, 2k]$. As the NA increases, the support of the monochromatic TF spans a larger range in the Fourier domain, leading to more overlapping and a larger condition number. At zero NA, by contrast, every monochromatic TF collapse to a delta function, and the Fourier transform of the pixel-wise CSI signal can precisely recovers the spectrum vector exactly—the principle underlying classical Fourier Transform Infrared Spectroscopy [22]. In Fig. 6 (c), the phase spectrum reflects the sample-introduced relative PCOR $\Delta_s(k)$ at each corresponding wavenumber. The phase spectrum appears flat because $\Delta_s(k)$ is assumed to be angle-independent. If the relative PCOR is zero, all monochromatic interferometric signals are in phase, resulting in zero phase for each monochromatic TF. Finally, Eq. (30) of Ref. [17] shows that in an aberration-free CSI system with uniform pupil illumination the total energy of each TF magnitude is independent of wavenumber. The simulations in Fig. 6 (b) confirm this prediction: even though the TFs broaden with NA, the area under every curve remains constant across wavenumbers.

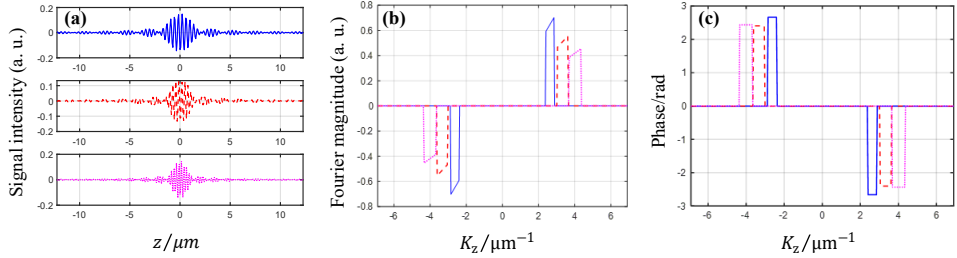

Fig. 6. Comparison of the column signals within the spectrum measurement matrix $\boldsymbol{A}$ and their

Fourier magnitudes at NA=0.55:(a) Column signals, (b) Fourier magnitudes.

In Fig. 7, the 1-D TF measurement matrix is shown (the phase matrix is not displayed), where each column represents the monochromatic TF at the corresponding wavenumber. This TF measurement matrix is completely equivalent to the spectrum measurement matrix, as shown in Fig. 3, as the Fourier matrix is a unitary matrix.

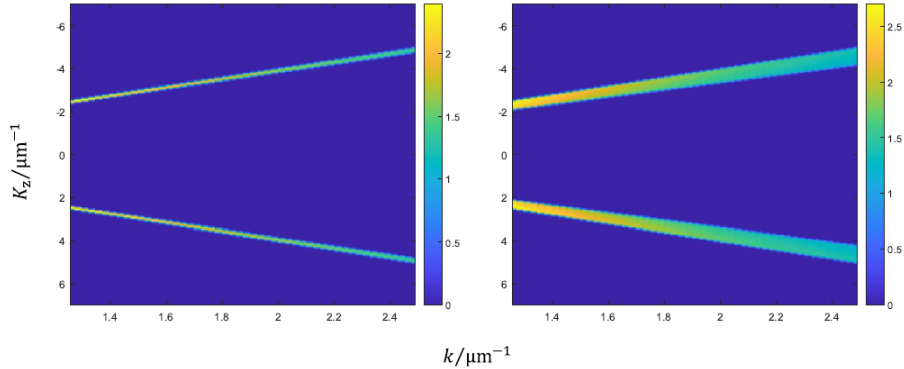

Fig. 7. 1-D TF measurement matrix:(a) NA=0.3, (b) NA=0.55.

## 3.2 Reflection measurement matrix

In Eq. (7), we only connect the spectrum and interferometric signal. To obtain the reflection spectrum, it is needed to remove the spectrum of the light source and the reference path. Thus, Eq. (7) can be reformulated to account for the spectral distribution of the light source as follows:

$$\boldsymbol{y} = \boldsymbol{A}_r |\boldsymbol{r}_s|, \tag{9}$$

where $\boldsymbol{A}_r$ denotes the reflection measurement matrix, given by $\boldsymbol{A}\text{diag}(\boldsymbol{s})\text{diag}(|\boldsymbol{r}_r|)$, and $|\boldsymbol{r}_s|$ denotes the measured sample's magnitude of the reflection spectrum. Here, $\text{diag}(\cdot)$ represents the operation of creating a diagonal matrix from a vector. For simplicity, the complex reflection coefficient spectrum of the reference path $|\boldsymbol{r}_r|$ is assumed to be 1 at each wavenumber.

Figure 8 shows the normalized singular values of both the spectrum measurement matrix $\boldsymbol{A}$ and the reflection spectrum measurement matrix $\boldsymbol{A}_r$ at NA=0.55. Since the smallest singular value of $\boldsymbol{A}_r$ is strictly positive, $\boldsymbol{A}_r$ is also full rank as $\boldsymbol{A}$. In principle, Eq. (9) can be solved using the least squares method to obtain the sample's reflection magnitude $|\boldsymbol{r}_s|$, assuming perfect knowledge of the source spectrum-which includes the spectral response of the light source and the monochrome camera. However, the smallest singular value of $\boldsymbol{A}_r$ is about two orders of magnitude smaller than that of $\boldsymbol{A}$, indicating that Eq. (9) is much more sensitive to perturbations and noise.

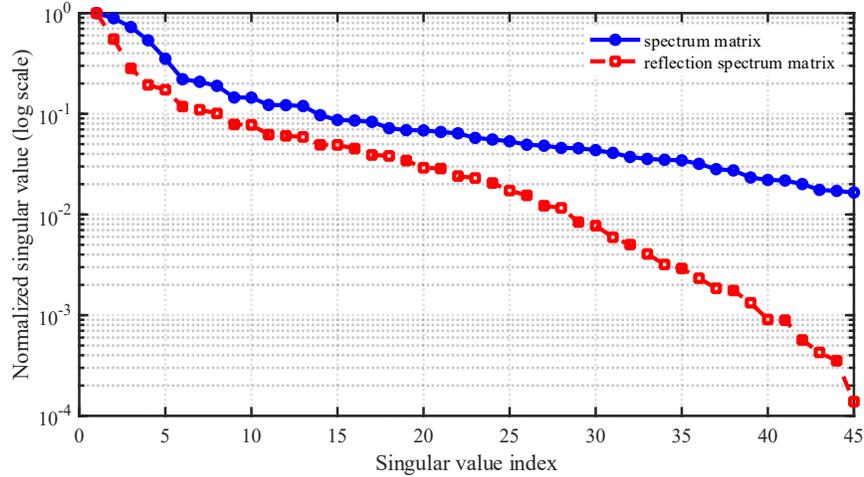

Fig. 8. Normalized singular values of the spectrum measurement matrix and reflection spectrum measurement matrix at NA=0.55.

Since the spectral distribution of the light source $\boldsymbol{s}$ often exhibits low intensities at certain wavenumbers, the reflection measurement matrix, $\boldsymbol{A}_r$ is truncated by retaining only those columns whose vector norms exceed 5% of the maximum column norm. Figure 9 illustrates the condition numbers of both the spectrum measurement matrix $\boldsymbol{A}$ and the reflection measurement matrix $\boldsymbol{A}_r$ across different NAs. It is evident that the condition number of $\boldsymbol{A}_r$ is approximately five times greater than that of $\boldsymbol{A}$, primarily due to the spectral distribution of the light source. In practical scenarios, various sources of error introduce perturbations in both the created measurement matrix $\boldsymbol{A}_r$ and the observed data $\boldsymbol{y}$. The larger condition number of the truncated reflection measurement matrix can significantly degrade the reconstruction performance of the sample's reflection magnitude $|\boldsymbol{r}_s|$.

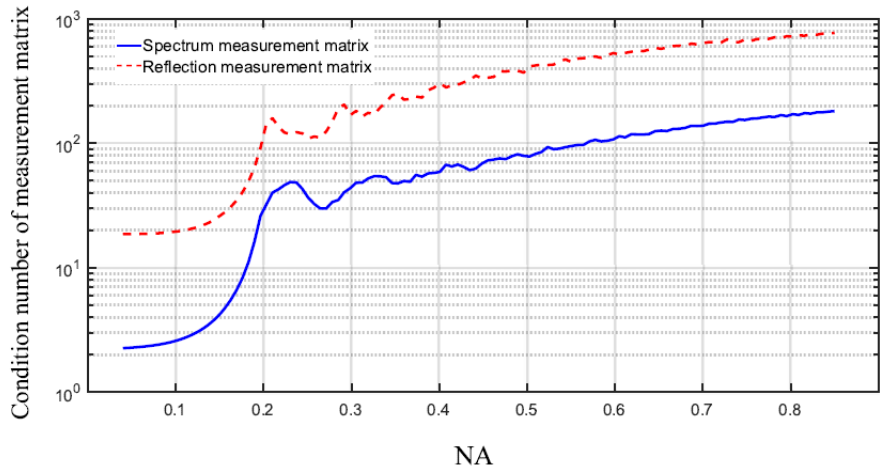

Fig. 9. Condition number of the measurement matrices at different NAs: (a) spectrum measurement matrix and (b) truncated reflection measurement matrix.

### 3.3 Conditioned reflection measurement matrix

To enhance the numerical stability of solving of Eq. (9), the reflection measurement matrix $A_r$ can be improved by modifying its structures. If the reflection magnitude spectrum varies slowly with respect to the wavenumber, the $n$ wavenumber points can be grouped into $n_c$ wavenumber channels, with each channel representing the combined contribution of approximately $n/n_c$ wavenumbers. Under this formulation, Eq. (9) can be reformulated as:

$$y = A_c r_c ,\qquad(10)$$

where the conditioned reflection measurement matrix $A_c$ has $n_c$ columns, and each column is constructed by summing the corresponding columns of $A_r$ within the associated wavenumber channel:

$$A_c(:,j_0) = \sum_{i \in \text{channel}_{j_0}} A_r(:,i) ,\qquad(11)$$

here, $j_0 \in \{1, \ldots, n_c\}$, and the index $i$ ranges from $(j_0 - 1) \cdot n/n_c + 1$ to $j_0 \cdot n/n_c$. The vector $r_c$ denotes the sample's reflection magnitudes at the center wavenumbers of the $n_c$ channels. Figure 10 illustrates the conditioning of the reflection measurement matrix $A_r$, demonstrating a trade-off between reduced spectral resolution and improved numerical stability.

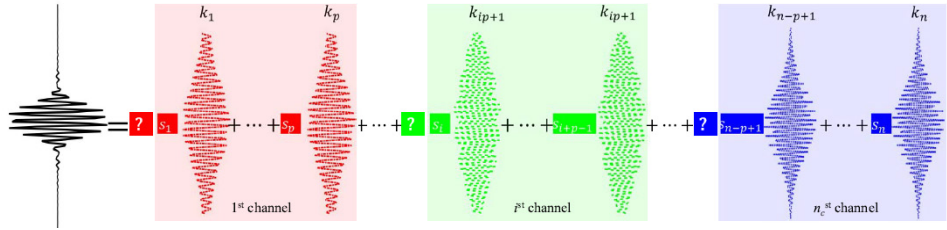

Fig. 10. Linear conditioned representation of a CSI signal.

The conditioning operation not only reduces the impact of low-intensity regions in the light source spectrum on the reconstruction of the sample's reflection coefficient, but also provides flexibility to adjust the number of wavenumber channels based on practical measurement requirements. For instance, when the numerical aperture (NA) is relatively low (e.g., below 0.3), the condition number of the conditioned measurement matrix is significantly smaller compared to cases with higher NA values (e.g., 0.55), as illustrated in Fig. 11. In such scenarios, the number of wavenumber channels can be increased to improve spectral resolution. Conversely, when the measured CSI signals exhibit high noise levels, reducing the number of channels can enhance the robustness of the reconstruction.

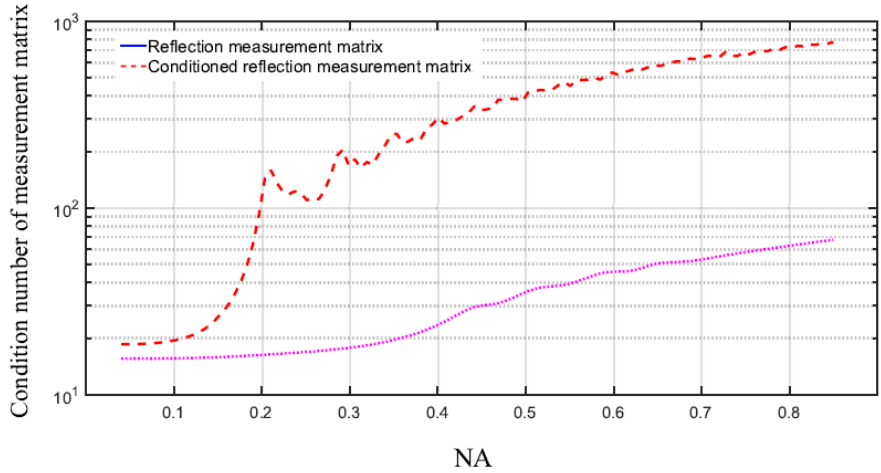

Fig. 11. Condition number of the measurement matrices at different NAs.

## 4. Quantitative analysis of the reflection magnitude spectrum reconstruction errors

Figure 12 illustrates the simulation and processing workflow for the CSI signal. The simulation process begins with setting the optical parameters of the CSI system—such as NA, the spectral distribution of the light source, and the scanning positions—all of which, unless otherwise specified, follow the settings described in Section 3.1. The sample surface height is randomly assigned, and the complex reflection coefficients for different metallic materials (including Au, Ag, and Cu) are calculated using the Fresnel equations. Based on these conditions, a series of CSI signals are simulated. To mimic real-world measurement noise, 40 dB of Gaussian noise is added to the signals.

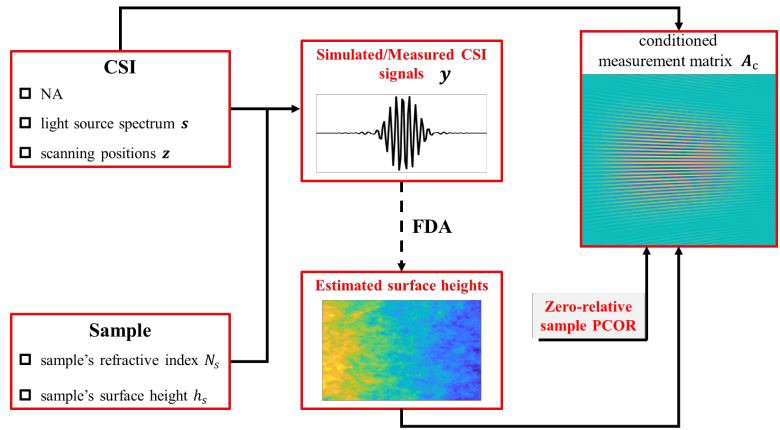

Fig. 12. Signal simulation and processing workflow for CSI signals.

For signal processing, Fourier domain analysis [4] is used to estimate sample surface heights from both simulated and measured signals. The spectrum measurement matrix $A$ is then constructed according to Eq. (8), using the known spectral distribution of light source and estimated surface heights. Notably, zero relative sample-induced PCOR is assumed. The spectral bandwidth is subsequently divided into 9 wavenumber channels, yield a conditioned reflection measurement matrix. Finally, the reflection magnitude spectrum is reconstructed by solving Eq. (10).

*4.1 Error propagation with condition number analysis*

In this section, we quantitatively analyze the error sources contributing to the discrepancy between the true values and the reconstructed reflection magnitude spectrum, using condition number analysis. According to linear algebra theory, the sensitivity of the solution $x$ in a linear system $Mx = b$ to perturbations in $M$ and $b$ can be approximated by:

$$\frac{\|\triangle x\|}{\|x\|} \approx \kappa(M) \frac{\|\Delta M\|_F}{\|M\|_F}, \tag{12}$$

and

$$\frac{\|\triangle x\|}{\|x\|} \approx \kappa(M) \frac{\|\Delta b\|_F}{\|b\|_F}, \tag{13}$$

where $\kappa(M)$ is the condition number of the measurement matrix $M$, $\triangle x$ denotes the perturbation of the reconstructed value of $x$ and $\Delta M$ and $\Delta b$ denote the perturbations in the matrix $M$ and observation vector $b$, respectively

Here, we use the term Relative Frobenius error (RFE) to quantitatively describe the perturbation of the measurement matrix:

$$\text{RFE}_M = \frac{\|\Delta M\|_F}{\|M_{real}\|_F} = \frac{\|M_{real} - M_{ideal}\|_F}{\|M_{real}\|_F}, \tag{14}$$

where $M_{real}$ represents the measurement matrix obtained from real-world measurements, and $M_{ideal}$ denotes the ideal, theoretical, or reference measurement matrix, derived under perfect conditions. Similarly, a RFE term is defined to quantify the perturbation of the observation vector:

$$\text{RFE}_b = \frac{\|\Delta b\|_2}{\|b\|_2}, \tag{15}$$

In this manuscript, the measurement matrix $M$ may refer to the spectrum measurement matrix $A$, the reflection measurement matrix $A_r$ and the conditioned reflection measurement matrix $A_c$. In the following sections, we focus primarily on the conditioned reflection measurement matrix $A_c$.

Numerous sources of error collectively influence both the captured interferometric signal $y$ and the measurement matrix. Inherent noise sources—such as camera noise, scanner vibrations, and environmental (ground) vibrations—directly impact the observed signal $y$. Additionally, several system-level imperfections contribute to deviations in the measurement matrix, including pupil apodization (e.g., non-uniform intensity distribution in the objective's pupil plane), discrepancies between the nominal and actual NA, scanner nonlinearity, inaccuracies in the spectral distribution of the light source (including the camera's spectral response), optical aberrations (particularly chromatic aberration), and the measured sample's PCOR [23-29], These factors often interact in complex ways, leading to deviations between the constructed measurement matrix $A_c$ and the true encountered in practical applications.

### 4.1.1 Camera noise

In CSI, camera noise is a significant inherent error source. In Fig. 13, the variation of $RFE_b$ with the signal-to-noise ratio (SNR) at an NA of 0.55 is illustrated. It is evident that $RFE_b$ decreases as the SNR increases. Notably, under common signal acquisition conditions with SNR values ranging from 30dB to 40dB, $RFE_b$ remains between approximately 0.01 and 0.03.

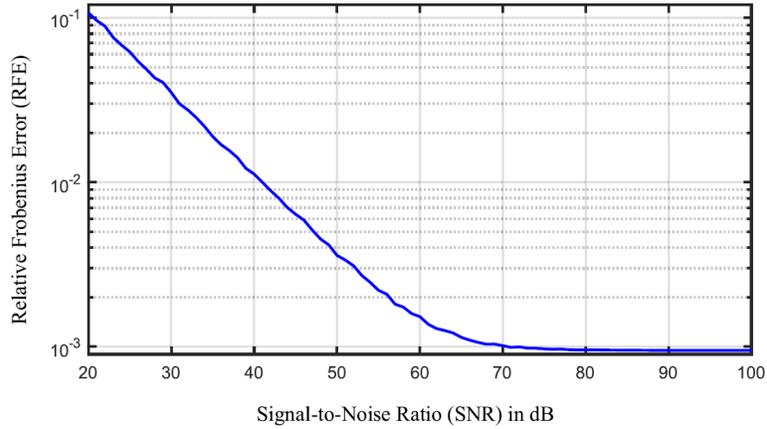

Fig. 13. Variation of $RFE_b$ with SNR at NA=0.55.

### 4.1.2 Inaccurate NA

Although the microscope objective (MO) has a nominal NA value, the actual NA can deviate from this nominal value. For instance, if we are using an MO with a nominal NA of 0.55, the real NA could deviate from the nominal value. Assuming the real NA is 0.56, the deviation from the nominal value is approximately 1.8%. In Fig. 14, the conditioned reflection measurement matrices are illustrated for NA=0.55, NA=0.56, and their difference. The $RFE_M$ between the ideal measurement matrix and the real measurement matrix could be as large as 0.07.

It should be noted that in the analysis of NA, all other variables are assumed to be perfectly accurate. This control variable method will be applied in subsequent analyses unless explicitly stated otherwise.

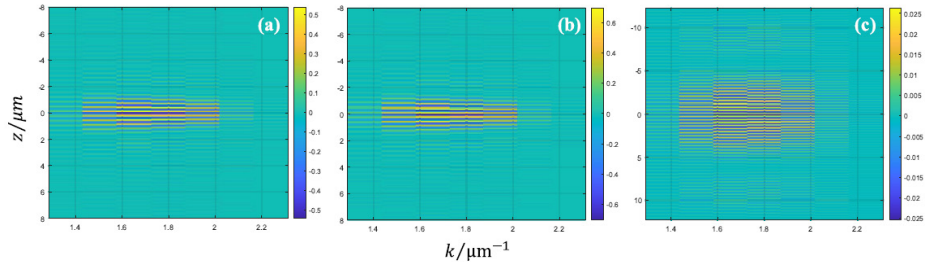

Fig. 14. Conditioned reflection measurement matrix: (a) NA=0.55, (b) NA= 0.56, (c) Difference.

*4.1.3 Pupil apodization*

The MO is typically assumed to have a uniform intensity distribution in the back-focal pupil plane. However, in practical applications, the pupil intensity distribution is often non-uniform. Assuming the non-uniform pupil intensity distribution maintains rotational symmetry, for example, the intensity decreases linearly by 5% from the center to the edge of the pupil, this non-uniformity must be accounted for. In Fig. 15, the conditioned reflection measurement matrices are illustrated for both uniform and non-uniform pupil intensity distributions, along with their difference. The $RFE_M$ between the ideal measurement matrix and the real measurement matrix can be as large as 0.035.

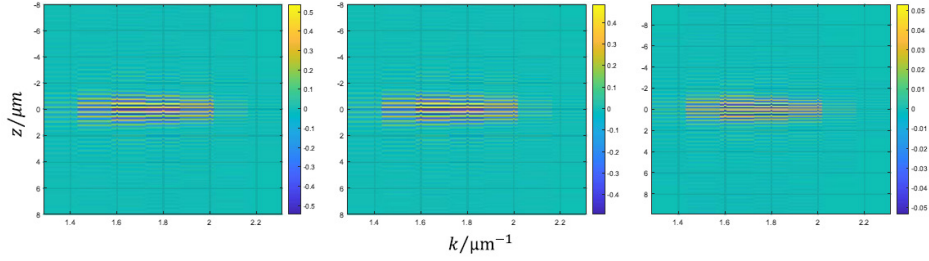

Fig. 15. Conditioned reflection measurement matrix: (a) Uniform intensity distribution in the pupil plane, (b) Non-uniform intensity distribution in the pupil plane, (c) Difference.

*4.1.4 Height reconstruction error*

From Eq. 8, it is evident that the sample surface height is a crucial factor in the creation of the spectrum measurement matrix, and thus in the conditioned reflection measurement matrix. Assuming that there are no other errors except for the sample surface height reconstruction error—which may arise from the Fourier analysis of the measured CSI signal—the $RFE_M$ between the ideal and real measurement matrices is illustrated in Fig. 16 for different sample surface height reconstruction errors.

The results clearly indicate that the $RFE_M$ increases sharply with larger height reconstruction errors. To maintain the $RFE_M$ below 0.05, the sample height reconstruction error should be kept under 2.5 nm.

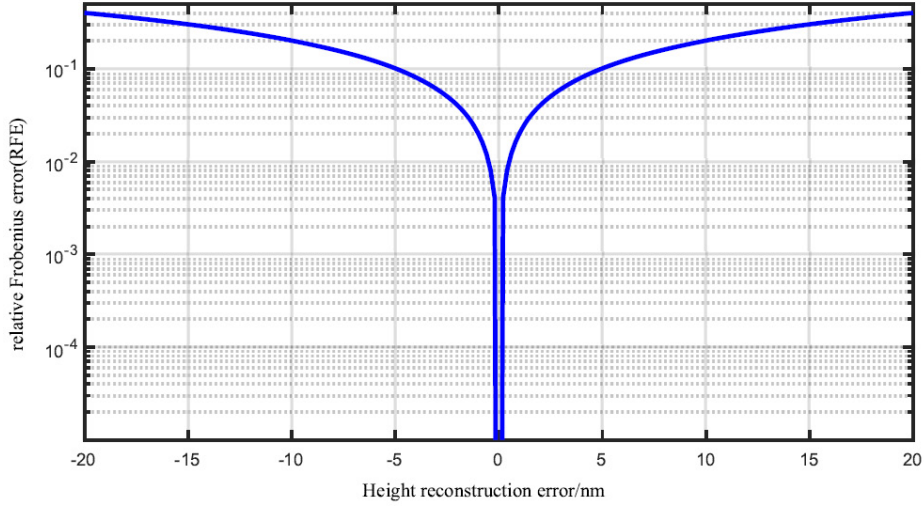

Fig. 16. Relative Frobenius error $RFE_M$ of the conditioned reflection measurement matrix with height reconstruction error.

### 4.1.5 Spectral errors of the light source

Accurate information about the spectral distribution of the light source is essential for creating the reflection measurement matrix, and thus the conditioned reflection measurement matrix. Typically, the spectral distribution obtained from a spectrometer is used. However, in this case, the spectral response of the monochrome camera is neglected. In Fig. 17, the spectral distribution of the light source is shown both with and without considering the camera response. The "used" spectrum neglects the camera response, while the "truth" spectrum accounts for it.

In Fig. 18, the conditioned reflection measurement matrices are illustrated using the spectral distribution of the light source without and with considering the camera response, and their difference. The $RFE_M$ between the ideal measurement matrix and the real measurement matrix can be as large as 0.08

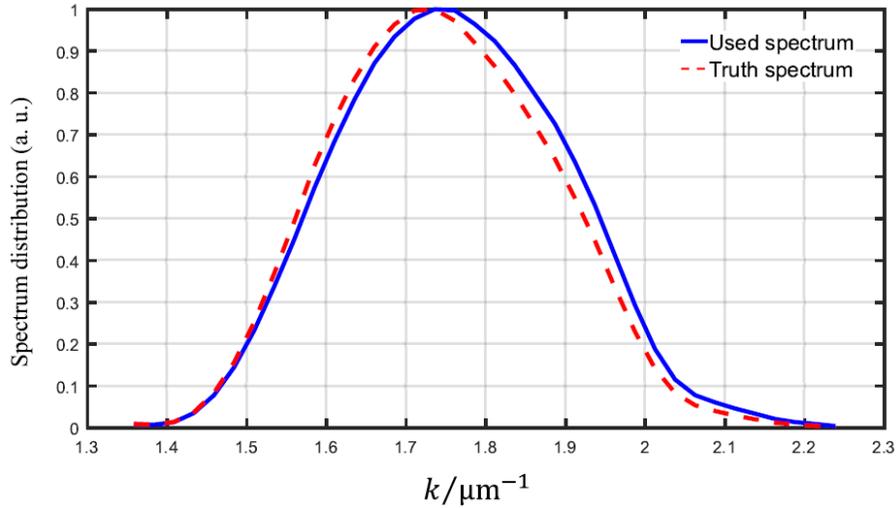

Fig. 17. Comparison of used spectral distribution of the light source, and truth spectrum.

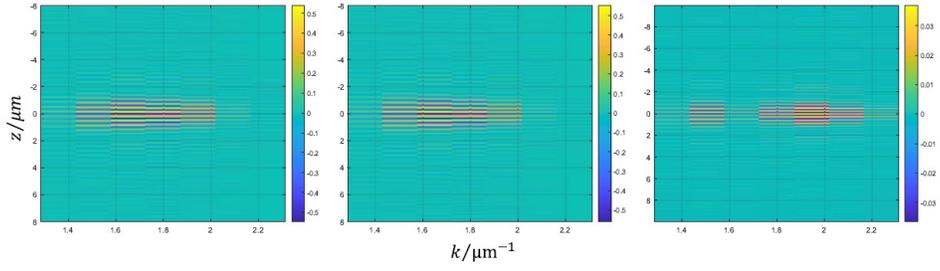

Fig. 18. Conditioned reflection measurement matrix: (a) Used spectral distribution of the light source, (b) Truth spectrum of the light source, (c) Difference.

### 4.1.6 Sample's PCOR

In the signal processing flow, the material's PCOR is usually not known in advance. Therefore, the generation of the spectrum measurement matrix often proceeds without considering the PCOR. Specifically, the material's PCOR, $\Delta_s$, is assumed to be zero. This assumption leads to a systematic error in the creation of the conditioned reflection measurement matrix. In Fig. 19, the conditioned reflection measurement matrices are illustrated both without and with consideration of the sample's PCOR, along with their differences. The $RFE_M$ between the ideal measurement matrix and the real measurement matrix can be as large as 1.8647. In other words, the ideal measurement matrix is totally different from the actual measurement matrix.

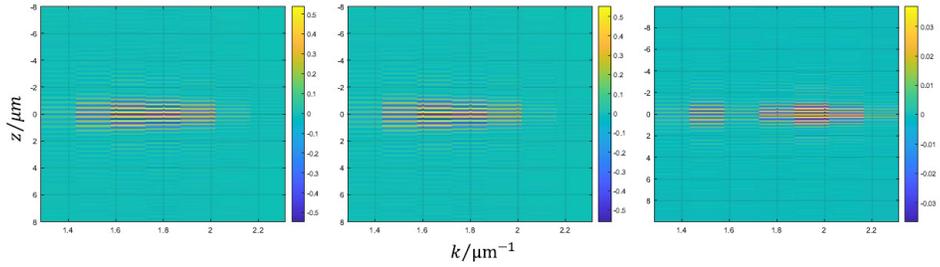

Fig. 19. Conditioned reflection measurement matrix: (a) Without sample's PCOR, (b) With sample's PCOR, (c) Difference.

In the signal processing workflow as in Fig. 12, the material's PCOR can result in sample surface height reconstruction errors [27]. These height reconstruction errors mitigate the effect of the PCOR in generating vastly different measurement matrices. In Fig. 20, the conditioned reflection measurement matrices are shown without considering the sample's PCOR but using the reconstructed height, with the sample's PCOR taken into account but using the ideal height, and the difference between the two. The RFE between the ideal and real measurement matrices can be as large as 0.24, which is significantly smaller than in the case where the ideal height is used while the PCOR is neglected.

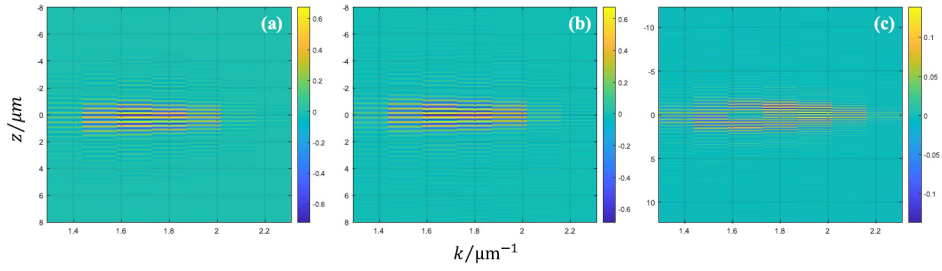

Fig. 20. Conditioned reflection measurement matrix: (a) Without sample's PCOR, (b) With sample's PCOR, (c) Difference.

Based on the above results, the dominant factor appears to be the material's PCOR. This effect causes phase misalignment in the monochromatic CSI signal at different wavenumbers, and while this is partially mitigated by the height reconstruction errors derived from the analysis of the CSI signals, the residual phases due to the sample's PCOR remain the primary influence. Nevertheless, the condition number of the conditioned reflection measurement matrix is about 40 at NA=0.55, indicating that relatively small errors—such as camera noise, inaccurate NA, pupil apodization, and spectral errors of the light source—could also significantly affect the accurate reconstruction of the reflection spectrum.

Furthermore, the chromatic dispersion of the optical system exhibits a similar effect to the sample's PCOR. In practical experimental scenarios, all of these errors may occur simultaneously, with some errors potentially canceling each other out or amplifying one another. Therefore, it is crucial to minimize all sources of error.

*4.2 Discussion*

At present, some of these errors can be calibrated and mitigated. The inherent random errors cannot be fully eliminated without compromising measurement time or incurring the cost of more expensive hardware. Typically, these errors can be mitigated through the use of the time-averaging technique [30]. The pupil apodization can be calibrated with a specialized Bertrand tube-lens [31], and the optical aberrations can be estimated by calibrating the 3-D monochromatic TFs at different wavenumbers with a near-perfect micro-sphere [32]. The scanner nonlinearity can be measured with a laser displacement interferometer [33]. It is noticeable that the sample's PCOR is not directly measurable in CSI and varies between different samples. Accurate determination of a sample's PCOR requires further calibration using additional instruments. To date, no effective method has been proposed to extract PCOR information in a conventional CSI.

## 5. Conclusions

we have developed a generalized mathematical framework that conceptualizes the CSI signal as a matrix multiplication of the spectrum measurement matrix with the element-wise product of the reflection and light source spectra, assuming spectral independence from incident angles.

Utilizing the light source spectrum and wavenumber channel summation, we employ a reflection spectrum measurement matrix and its conditioned counterpart to linearly relate the CSI signal to the absolute reflection spectrum. Through extensive simulations, we have quantitatively analyzed the impacts of various error sources, including NA inaccuracy, pupil apodization, height reconstruction errors, spectral inaccuracies of the light source, and PCOR from the sample. Our findings reveal that PCOR is the dominant factor influencing the accurate reconstruction of the reflection coefficient spectrum.

This linear algebraic representation simplifies the traditional model while accommodating both spatial and spectral coherence. It allows us to identify and quantify the error sources that significantly affect the reconstruction process. By characterizing the CSI using several measurement matrices, we have provided a robust framework that enhances the understanding and application of CSI in complex imaging scenarios.

## Funding

## Acknowledgment

The authors thank xx for insightful discussions on xx. Also, xx.

## Disclosures

The authors declare that there are no conflicts of interest related to this article.